\documentclass[12pt,preprint]{aastex}

\slugcomment{ApJ in press}

\shorttitle{Modeling the Afterglow of GRB 030329}
\shortauthors{Huang Y.F. et al.}

\begin{document}


\title{Modeling the Optical Afterglow of GRB 030329}


\author{Y. F. Huang}
\affil{Department of Astronomy, Nanjing University, Nanjing 210093, China}
\affil{Department of Physics, The University of Hong Kong, 
         Pokfulam Road, Hong Kong, China}
\email{hyf@nju.edu.cn}

\author{K. S. Cheng}
\affil{Department of Physics, The University of Hong Kong, 
         Pokfulam Road, Hong Kong, China}

\and

\author{T. T. Gao}
\affil{Department of Astronomy, Nanjing University, Nanjing 210093, China}




\begin{abstract}
The best-sampled afterglow light curves are available for GRB 030329. 
A distinguishing feature of this event is the obvious rebrightening 
at around 1.6 days after the burst. Proposed explanations for the
rebrightening mainly include the two-component jet model and the refreshed 
shock model, although a sudden density-jump in the circumburst environment 
is also a potential choice. Here we re-examine the optical afterglow of 
GRB 030329 numerically in light of the three models. In the density-jump 
model, no obvious rebrightening can be produced at the jump moment. 
Additionally, after the density jump, the predicted flux density 
decreases rapidly to a level that is significantly below observations.   
A simple density-jump model thus can be excluded. In the two-component 
jet model, although the observed late afterglow (after 1.6 days)
can potentially be explained as emission from the wide-component, 
the emergence of this emission actually is too slow and it does not 
manifest as a rebrightening as previously expected. The energy-injection
model seems to be the most preferred choice. By engaging a sequence of
energy-injection events, it provides an acceptable fit to the rebrightening 
at $\sim 1.6$ d, as well as the whole observed light curve that extends to 
$\sim 80$ d. Further studies on these multiple energy-injection processes 
may provide a valuable insight into the nature of the central engines of 
gamma-ray bursts. 
\end{abstract}





\section{Introduction}

GRB 030329, with a fluence as large as $\sim 1.18 \times 10^{-4}$ ergs/cm$^2$ 
(Ricker et al. 2003; Vanderspek et al. 2004) and 
being in the top 1\% of all detected gamma-ray bursts, 
is a watershed event in the field. Lying at a redshift of $z=0.1685$ 
(Greiner et al. 2003a), it is the closest classical gamma-ray burst (GRB) 
to date. For the first time, an unambiguous underlying 
type Ic supernova was revealed spectroscopically about one week after 
the trigger (Hjorth et al. 2003; Stanek et al. 2003; Matheson et al. 2003). 
This already suspected connection between GRBs and core-collapse supernovae,
which was first hinted at by the amazing coincidence of GRB 980425 and 
SN 1998bw (Galama et al. 1998), is now firmly established, finally 
shedding light on the previously unclear nature of long GRBs. 
Additionally, the radio afterglow of GRB 030329 was resolved with Very 
Long Baseline Interferometry observations, leading to a direct measurement 
of the size of a cosmological GRB remnant for the first time (Taylor et 
al. 2004, 2005). The observed expansion rate of the remnant, being 
generally consistent with theoretical expectations, provides valuable 
evidence for the standard fireball model (Oren, Nakar, \& Piran 2004; 
Granot, Ramirez-Ruiz, \& Loeb 2004). Furthermore, a polarization
light curve of unprecedented detail was obtained (Greiner et al. 2003b). Observed
polarization, with significant variability, is at the percent level, 
casting light on the structure of the jet, the configuration of
internal magnetic field, and other micro-physics of the blastwave.

So far, GRB 030329 is also the event with the most copious afterglow 
data, due to its extremely bright afterglow. 
A very detailed R-band afterglow light curve has been compiled 
by Lipkin et al. (2004). The available R-band data spans from $ \sim $ 0.05 d
to $\sim$ 80 d, with a total of 1644 points, which is unprecedented. 
The R-band light curve shows many interesting features. First, an 
obvious bending appears at $t \sim 0.5$ d, which can be satisfactorily 
interpreted as a jet break (Uemura et al. 2003). Secondly, the afterglow rebrightened 
significantly and rapidly at $t \sim 1.6$ d. Thirdly, obvious variability 
has also been observed during $t \sim 2.3$ --- 7 d. Finally, the afterglow 
rebrightened markedly again at $t > 20$ d as compared with the simple power-law 
extrapolation, which in fact reflects the contribution from the 
underlying supernova, emerging as the GRB afterglow itself fades away. 
The copious observations and the interesting afterglow behavior have 
made GRB 030329 an amazing example, attracting the attention of many 
authors. 

Berger et al. (2003) suggested that the rebrightening at $t \sim 1.6$ d
can be explained by adopting a two-component jet model. In their 
framework, the central, narrower, faster jet can account for the 
light curve break at $t \sim 0.5$ d, while the outer, wider, slower 
jet, which intrinsically carries more kinetic energy, will finally 
outshine the former and naturally give birth to the observed rebrightening 
at $\sim 1.6$ day. On the other hand, Granot, Nakar, \& Piran (2003) 
suggested another model for the rebrightening. They proposed 
that a refreshed shock, i.e., energy injected into the blastwave by
an additional shell from the central engine, can boost the brightness.
Only simplified analytical approaches have been devoted 
to this important question until now. It is thus worthwhile to revisit 
the issue by carrying out realistic and more accurate numerical calculations. 

In this study, we will model the R-band afterglow light curve of 
GRB 030329 numerically, paying special attention to the rebrightening 
at $t \sim 1.6$ d. We base our calculations on three candidate models, 
i.e., the density-jump model, the two-component jet model, 
and the energy-injection model. Our
paper is organized as follows. We first describe the details of our calculations,
including the dynamics and the radiation process in \S 2. We then 
examine the observed R-band light curve in the framework 
of the three models respectively in \S 3. We discuss our results 
and present our conclusions in \S 4. 

\section{Dynamics and Radiation Process}

In the standard fireball model, afterglows are produced when the fireball,
either isotropic or collimated, ploughs through the circumburst medium,
producing a strong blastwave that accelerates swept-up electrons (For 
recent reviews, see van Paradijs, Kouveliotou, \& Wijers 2000; 
M\'esz\'aros 2002; Piran 2004; Zhang \& M\'esz\'aros 2004). Afterglows 
are observed when synchrotron  photons are emitted by these accelerated 
electrons (Sari, Piran, \& Narayan 1998), although inverse Compton 
scattering may also play a role in some cases (Wei \& Lu 2000a; Sari \&
Esin 2001). The conditions involved in GRB afterglows are complicated. 
For example, 
the blastwave may be either highly radiative or highly adiabatic, and may 
experience the ultra-relativistic phase and the Newtonian phase 
sequentially. In case of jets, the remnant may expand laterally or not.
The circumburst medium may be either homogeneous or wind-like. 
The shock-accelerated electrons may be adiabatic or cool in real time. 
The final afterglow light curve also strongly depends on the frequency 
that we are observing at. 
Simple analytical results are available for the whole process of the afterglow, 
but detailed expressions can be given only when the conditions involved are 
highly simplified (Zhang \& M\'esz\'aros 2004). 

On the other hand, there are also some factors that cannot be
easily incorporated into analytical considerations, among them is the 
 equal arrival time surface effect (Waxman 1997; Sari 1997; 
Panaitescu \& M\'esz\'aros 1998). This ingredient will definitely affect 
the smoothness and variability of GRB afterglow light curves significantly. 
Although analytic expressions for equal arrival time surfaces can be 
derived under some simplified assumptions (Bianco \& Ruffini 2005), their
exact effects on the light curve still cannot be included in usual 
analytical expressions. Numerical evaluation will be the only efficient 
solution in some circumstances, especially when rapid variability is involved. 

A simple model that can be applied under various conditions 
addressed above, and which is also very convenient to solve numerically,
has been developed by Huang et al. (1999, 2000a, 2000b; Huang \& Cheng 2003). 
We will use this model for the current study. In this model, the evolution of the 
bulk Lorentz factor ($\gamma$) of the shock-accelerated circumburst medium
is given by (Huang, Dai, \& Lu 1999),
\begin{equation}
\label{dgdm1}
\frac{d \gamma}{d m} = - \frac{\gamma^2 - 1}
       {M_{\rm ej} + \epsilon m + 2 ( 1 - \epsilon) \gamma m}, 
\end{equation}
where $m$ is the mass of swept-up medium and $M_{\rm ej}$ is the initial 
mass of the fireball. $\epsilon$ is the radiative efficiency, which
equals 1 for a highly radiative blastwave, and equals 0 in the adiabatic 
case. Equation~(1) has the virtue of being applicable in both the
ultra-relativistic and the non-relativistic phases (Huang et al. 1999). 
For collimated outflows, the lateral expansion is realistically 
described by (Huang et al. 2000a, 2000b),
\begin{equation}
\label{dthdt2}
\frac{d \theta}{d t} = \frac{c_{\rm s} (\gamma + \sqrt{\gamma^2 - 1})}{R},
\end{equation}
with the comoving sound speed $c_{\rm s}$ given by
\begin{equation}
\label{cs3}
c_{\rm s}^2 = \hat{\gamma} (\hat{\gamma} - 1) (\gamma - 1) 
	      \frac{1}{1 + \hat{\gamma}(\gamma - 1)} c^2 , 
\end{equation}
where $\theta$ is the half-opening angle, $R$ is the radius, and 
$\hat{\gamma} \approx (4 \gamma + 1)/(3 \gamma)$ is the adiabatic 
index.

To calculate synchrotron radiation from shock-accelerated electrons, 
a realistic electron distribution function (Dai, Huang, \& Lu 1999; 
Huang \& Cheng 2003) that takes into account the cooling effect 
(Sari, Piran, \& Narayan 1998) will be adopted. Especially, since 
we will assume in our calculations a value smaller than 2 for the 
electron power-law distribution index, $p$, the minimum Lorentz 
factor of electrons should be given by 
\begin{equation}
\gamma_{\rm e,min}=\left[\left(\frac{2-p}{p-1}\right)\left(\frac{m_p}{m_e}\right)
    \epsilon_e (\gamma-1) (\gamma_{\rm e,max}-1)^{p-2}\right]^{1/(p-1)} +1,
    \;\; (1<p<2),
\end{equation}
where $m_{\rm p}$ and $m_{\rm e}$ are masses of proton and electron 
respectively, $\gamma_{\rm e,max}$ = $10^8 (B'/1 {\rm G})^{-1/2}$ is 
the maximum Lorentz factor of electrons, with $B'$ being the comoving
magnetic field strength. Equation~(4), which slightly differs from the 
expression given by Dai \& Cheng (2001) for electrons with a flat 
spectra, is more general since it is applicable even in the deep 
Newtonian phase, when $\gamma_{\rm e,min}$ is less than a few and 
most electrons are no longer ultra-relativistic (Huang \& Cheng 2003). 

\section{Numerical Results}

In this section we study the optical afterglow of GRB 030329 numerically, 
paying special attention to its rebrightening at $t \sim 1.6$ 
d. We take the R-band light curve provided by Lipkin et al. (2004) as the 
observed template, which has the advantage of having the widest time-span,
the most prolific data points, and also the least systematic discrepancy. 
However, the original data of Lipkin et al. includes contribution from the 
host galaxy and the underlying supernova. The host galaxy magnitude is 
$R = 22.66$ (Gorosabel et al. 2005). Using the observed light curve of SN 
1998bw as a template (Galama et al. 1998; Zeh, Klose, \& Hartmann 2004), 
the brightness of the supernova has also been determined by Zeh, Klose, \& 
Hartmann (2005). 
A pure R-band afterglow light curve is thus available for GRB 030329 after 
subtracting these extra components and correcting for Galactic
extinction (according to Schlegel, Finkbeiner, \& Davis 1998). Here we use
the pure afterglow light curve as the final template. We will try to fit 
it in light of three detailed models which all have the potential of 
producing the rebrightening: the density-jump model, the two-component 
jet model, and the energy-injection model.

\subsection{Density-Jump Model}

A possible model that can potentially produce a rebrightening in  
GRB afterglows is the 
so called density-jump model. Analytically it has been shown that when the
blastwave encounters a sudden density increase in the medium, the afterglow
emission will be enhanced temporarily (Lazzati et al. 2002; Nakar \& Piran 
2003; Dai \& Wu 2003; Tam et al. 2005). We have examined GRB 030329 in this 
framework. The jet involved is assumed to have an initial half-opening angle of
$\theta_{\rm N} =0.05$, with an isotropic kinetic energy $E_{\rm iso} = 
3.5 \times 10^{53}$ ergs, and an initial Lorentz factor $\gamma_{\rm 0} 
=300$. Other parameters are taken as: electron energy ratio 
$\epsilon_{\rm e} = 0.1$, magnetic energy ratio $\epsilon_{\rm B} = 0.01$, 
electron spectral index $p = 1.9$, and the luminosity distance 
$D_{\rm L} = 0.8$ Gpc (this value is derived by assuming a cosmology 
of $H_0 = 71$ km/s/Mpc, $\Omega_{\rm M}= 0.27$, $\Omega_\Lambda = 0.73$). 
The number density of the circumburst medium is initially assumed to be
$n = 2$ /cm$^3$, but it increases abruptly by a factor of 10 or 100 at 
the observer's time $t = 1.1 \times 10^5$ s. The final results are 
illustrated in Figure~1. 

In our theoretical light curves
no obvious rebrightening can be seen. The analytically predicted temporary 
rebrightening is smeared out by the equal-arrival time surface effect. 
Additionally, well after the density-jump, the afterglow flux decreases 
more steeply. For a higher density contrast, the fading of the afterglow
is even more obvious. This trend is consistent with the results of Tam et al. 
(2005) for cylindrical jets. Our results are also consistent with the other 
authors' conclusion that density fluctuations are usually unable to produce
either a sharp variation or a steep increase in the light curve (Piran, 
Nakar, \& Granot 2003; Nakar \& Piran 2003). 

Figure~1 shows clearly 
that the density-jump model can not explain the basic feature of GRB 030329,
the rebrightening at 1.6 d. It even can not explain the observed emission 
at $t \geq 2 \times 10^5$ s. A simple density-jump 
explanation thus can be completely excluded for GRB 030329. 

\subsection{Two-Component Jet Model}

The simplest jet model involves a homogeneous conical outflow. However,
in reality jets can have complicated structures (M\'esz\'aros, Rees, 
\& Wijers 1998; Dai \& Gou 2001; Rossi, Lazzati, \& Rees 2002; Kumar 
\& Granot 2003; Salmonson 2003; B. Zhang et al. 2004; W. Zhang, Woosley,
\& Heger 2004). A two-component jet consists of two components: a
narrow ultra-relativistic outflow and a wide but mildly relativistic 
ejecta, which are usually assumed to be coaxial (Frail et al. 2000; 
Ramirez-Ruiz, Celotti, \& Rees 2002; Berger et al. 2003; Sheth et al. 
2003; Huang et al. 2004; Peng, K\"onigl \& Grant 2005; Wu et al. 2005).
It has been suggested that the two-component jet model can give a 
satisfactory explanation to the multiband observations of GRB 030329
(Berger et al. 2003; Sheth et al. 2003): the gamma-ray and early
afterglow emission come from the narrow component, while the 
radio and optical afterglows beyond $\sim 1.5$ days are produced 
by the wide component. The half-opening angles of the two components 
are even estimated as $\sim 5^{\rm o}$ and $\sim 17^{\rm o}$ respectively
(Berger et al. 2003). The total intrinsic kinetic energy of the whole
jet is perfectly consistent with the standard energy reservoir
hypothesis (Frail et al. 2001; Panaitescu \& Kumar 2001; Bloom, 
Frail, \& Kulkarni 2003). 

We have tried to fit the R-band light curve of GRB 030329 by using
the two-component jet model numerically. The best results are illustrated 
in Figure~2. In our calculations, we evaluate the parameters as follows. 
For the narrow component, the initial half-opening angle is 
$\theta_{\rm 0,N} =0.05$, isotropic kinetic energy $E_{\rm N,iso} = 
3.0 \times 10^{53}$ ergs, and initial Lorentz factor $\gamma_{\rm 0,N} 
=300$. For the wide component, these parameters are $\theta_{\rm 0,W} =0.15$,
$E_{\rm W,iso} = 3.0 \times 10^{53}$ ergs, $\gamma_{\rm 0,W} =30$ 
respectively. Other parameters that are common to the two components are:
$\epsilon_{\rm e} = 0.1$, $\epsilon_{\rm B} = 0.01$, $n = 2$ /cm$^3$, $p = 1.9$, 
and $D_{\rm L} = 0.8$ Gpc. Some of our parameters differs from those recommended 
by Berger et al. (2003; also see Friedman \& Bloom 2005).  For example, we need
an $E_{\rm N,iso}$ that is larger by about a factor of 10, since in our
numerical calculations we take into account the deceleration of the 
blastwave before the usual deceleration radius. Also our $\theta_{\rm 0,N}$ 
is smaller, since the lateral expansion plays a subtle role in the process. 

Figure $2a$ shows clearly that the narrow component can give an acceptable 
fit to the observed light curve when $t < 10^5$ s, also the wide component
emission can give a marginally acceptable explanation for observations of 
$t > 2 \times 10^5$ s. However, when the emission from the two components 
is added together, the final light curve is disappointingly too smooth 
at $10^5 {\rm s} < t < 2 \times 10^5$ s. In other words, 
the model cannot reproduce the observed rapid rebrightening at $ t \sim 1.6$ 
d. The key problem is that the wide component emission peaks at  
$\sim 4 \times 10^3$ s, too early as compared with observations. Additionally,
the dotted line in Figure~$2a$ is very smooth at around the peak, so that 
it obviously has no hope to account for the rapid variability even if the peak 
were properly postponed. 

The difficulty of a simple two-component jet model to explain the rapidness
of such a rebrightening has been realized by Huang et al. (2004) in an earlier
study. They went further to conjecture that some subtle details, such as the 
overlap effect of the two components, may help to relax the difficulty (note
that in the calculations of Huang et al. (2004), it was assumed that $t=0$
at the deceleration radius). In Fig~$2b$, we have re-calculated the 
theoretical light curve by assuming that the wide component is a hollow cone 
since its central portion is occupied by the narrow component. In this case, 
the peak of the wide component emission is significantly postponed as expected. 
However, the light curve becomes even smoother near the peak. This
modification is thus essentially of no help in accounting for the rapidness of 
the rebrightening. 

In fact, the failure of the two-component jet model to reproduce the rapid
rebrightening at $t \sim 1.6$ d is not a surprize. In Berger et al. 
(2003), we notice that the rebrightening is still not rapid enough even
in their idealized analysis. A similar trend can also be seen in a superceding
detailed study on the two-component jet model by Peng, K\"onigl, \& Granot
(2005). In our current study, the equal arrival time surface effect and the 
realistic dynamical transition at the deceleration radius add together to  
further suppress the variability. Additionally, if a more complex dynamical
model as suggested by Granot et al. (2002) is adopted, things will surely
get even worse. In short, although the two-component jet model can give 
a feasible explanation for the overall R-band light curve, it is not 
satisfactory in reproducing the rapid rebrightening at $t \sim 1.6$ d. 
However, given that the existence of two jets has been clearly indicated 
by two breaks (at $\sim 0.5$ d and $\sim 5.5$ d respectively) in the 
observed light curve, the two-component jet model is still an attractive
idea for GRB 030329. We will further discuss some schemes that may 
ameliorate this idea in the final section of our paper. 

\subsection{Energy-Injection Model}

Although typical long GRBs last for only tens of seconds, the central 
engine can actually be active for much longer, supplying energy into
the blastwave during the afterglow phase. This can naturally lead to 
the rebrightening of GRB afterglows. Evidence for such activities has 
been found in a few events (Piro 1998; Dai \& Lu 1998, 2001; Zhang \& 
M\'esz\'aros 2001, 2002; Bjornsson et al. 2002; Bjornsson, Gudmundsson, 
\& Johannesson 2004; Burrows et al. 2005; King et al. 2005; Watson et
al. 2005; Cusumano et al. 2005).

Energy injection can be accomplished in various forms, on very different 
timescales. If the central engine is a rapidly rotating millisecond 
pulsar, a huge amount of rotation energy can be naturally injected into
the GRB remnant either in the form of a Poynting flux or a relativistic 
particle flux, when the rotating pulsar gradually brakes down (Dai \&
Lu 1998; Zhang \& M\'esz\'aros 2001). In this
case, the energy injection is a continuous process whose timescale is 
determined by the braking mechanism. Another possibility is that, since
the standard fireball model of GRBs resorts to internal shocks to 
produce the observed highly variable $\gamma$-ray light curve in the 
main burst phase, it is very likely that the central engine may also
give birth to some late slow shells, which catch up with the main 
remnant only in the afterglow phase (Rees \& M\'esz\'aros 1998; 
Kumar \& Piran 2000; Sari \& M\'esz\'aros 2000; Piran, Nakar, \&
Granot 2003; King et al. 2005). In this case, the energy supply will be 
completed relatively quickly, producing an essentially instantaneous 
energy injection. 

For the rebrightening of the afterglow of GRB 030329, energy-injection is
surely a potential explanation (Granot, Nakar, \& Piran 2003). 
Actually, Granot et al. (2003) 
suggested that in addition to the major energy injection occurring
at $t \sim 1.6$ d, there were furthermore three minor energy injection 
processes occurring at $t \sim 2.4$ d, 3.1 d, and 4.9 d respectively, 
giving birth to the observed subtle light curve variations
between  (2 --- 6) $\times 10^5$ s. 

We now fit the afterglow of GRB 030329 numerically by adopting the 
energy-injection model. Since the observed rebrightening at $t \sim 1.6$ 
d is so rapid, we believe that a quick energy-injection is necessary. 
In our calculation, we assume that an amount of kinetic energy that 
equals the initial energy ($E_0$) of the primary GRB ejecta is supplied into
the blastwave at the observer's time $t \sim 1.1 \times 10^5$ s. 
For simplicity, we assume that the energy supply is completed instantly. 
At $t = 4 \times 10^5$ s, we notice that another energy injection at the 
amplitude of $0.4 E_0$ is needed to account for the observed emission 
between $4 \times 10^5$ s and $1 \times 10^6$ s. This roughly corresponds
to the 4th energy injection process suggested by Granot et al. (2003). 
In Granot et al.'s study, the time span of the observed light curve is
$t < 9$ d. Here, when we expand the light curve to $t \sim 80$ d, we
find that an additional energy injection (with $0.6 E_0$) is necessary,
which occurs at about $1.2 \times 10^6$ s. Our final numerical results 
are shown in Figure~3. 

Interestingly enough, we find that the energy-injection at $t \sim 
1.1 \times 10^5$ s really can produce an obvious rebrightening as expected. 
The energy-injection model is thus better than the two previous models at 
least in this aspect. The relative residual of the solid line in Figure~3 
is generally less than 20\%, so that the overall fit can also be evaluated 
as acceptable. 

However, we also notice that there are still some obvious problems 
in the fit. First, the observed light curve shows a sharp jet break 
at $t \sim 0.5$ d ($\sim 4.3 \times 10^4$ s), but the theoretical 
light curve is simply too smooth, which leads to a systematic residual of 
$\sim$ 15\% during $2 \times 10^4$ s --- $5 \times 10^4$ s. 
It has been noted that a small half-opening angle of the jet 
can help to make the break sharper (Wei \& Lu 2000b; Huang 
et al. 2000a). In fact, in our current study, in order to get a break 
that is as sharp as possible, we have assumed a very small initial 
half-opening angle for the jet, i.e. $\theta_0 = 0.05$. Since the 
decay of the afterglow of a narrower jet is generally slightly faster,
we then have to assume a relatively flat spectrum for the shock-accelerated 
electrons ($p = 1.9$) so as to match the observed decay rate of 
$F_{\rm R} \propto t^{-0.85}$ before the jet break. However the 
theoretical break is still too shallow. In fact the same problem also 
exists in Figures~1 and 2. The sharpness 
of the observed light curve breaks actually is a general challenge to 
theorists, since numerical results by a few authors have shown 
that the predicted light curve break usually is not so sharp (Panaitescu 
\& M\'esz\'aros 1998; Moderski, Sikora, \& Bulik 2000; Wei \& Lu 
2000b). With plenty of observational data points before and after the 
jet break, GRB 030329 will be a valuable example that can be used to study 
the sharpness problem carefully. These studies may help to address many important 
issues of GRB afterglows, such as the initial opening angle of the jet, 
the effect of the lateral expansion, the influence of the equal arrival 
time surfaces, and so on. We thus suggest that the early afterglow  
of GRB 030329 ($t < 10^5$ s) deserves to be paid special attention to, and further 
detailed numerical study should be carried out. 

Secondly, although an obvious brightness enhancement is produced by the 
energy injection at $t \sim 1.6$ d, the theoretical rebrightening is 
still not rapid enough as compared with observations. As a result, 
we see that the relative residual reaches $\sim -20$\% before the 
rebrightening, and reaches $\sim +15$\% thereafter. Here we have 
already assumed an instantaneous energy injection. It is suspected 
that things might get even worse in reality since the energy injection
will surely take some time. However, at least two factors may help to
ease this unsatisfactory situation: (1) The energy injection itself may be a 
complicated process. For example, additional forward shocks or even 
reverse shocks may form when the slow shell collides with the original 
jet, and extra emission from these shocks may 
make the rebrightening more significant. But discussion of these 
extra emission will involve some largely uncertain conditions (such 
as the thickness, the composition, and the speed of the energy-injection 
shell), and will not be conducted here; (2) The half-opening angle of 
the injected shell may be another important factor. At the time of the 
energy injection, the opening angle of the original jet already 
increases to $\theta \approx 0.15$ due to lateral expansion. 
In our calculation, we have assumed that the energy 
is supplied to the whole jet homogeneously for simplicity. But it is  
probable that the injected shell itself may have an opening angle much
smaller than 0.15, then the energy supply will be restricted only to a small 
portion of the original jet, as already illustrated by Granot et al. (2003). 
In that case, the timescale of the rebrightening can be greatly reduced. 
Again, detailed consideration will involve some uncertain conditions,
such as the initial opening angle and the sideways expansion of the 
energy-injection shell. 

Thirdly, there are some subtle variations in the observed light curve
for $2 \times 10^5 {\rm s} \leq t \leq 8 \times 10^5$ s. As suggested 
by Granot et al. (2003), these variations may be due to further minor 
energy injections. In our calculations, we do not include the second and 
the third energy injection events proposed by Granot et al. In fact, since 
our modeling is still very coarse and highly simplified, we believe
that the observed fine structures will not be satisfactorily reproduced
even if all the minor energy injections are incorporated. To completely 
solve the problem, we may need to carefully consider the factors 
related to the second problem as addressed above. 

In short, GRB 030329 is a special but important event. Its afterglow 
behavior is very complicated and a satisfactory fit to the overall 
R-band light curve is not an easy task (Zeh, Klose, \& Kann 2005). 
However, after comparing all the three models examined in our current
study, we propose that the energy-injection model is the most appropriate 
one for GRB 030329, especially when the rebrightening at $t \sim 1.6$ d 
are taken into account. 

\section{Conclusion and Discussion}

The optical afterglow of GRB 030329, with a notable rebrightening at $t \sim 
1.6$ d, is re-examined numerically in light of three candidate models. 
In the density-jump model, no obvious rebrightening can be reproduced 
at the moment when the density increases abruptly. Additionally, the 
predicted flux density decreases significantly well after the density-jump, 
evidently in contrast with the observations. 
In the two-component jet model, emission from the wide component
can significantly boost the afterglow and thus can roughly fit the 
late afterglow of GRB 030329. However, the predicted rebrightening is 
still far too slow when compared with observations. In fact, no obvious 
bump can be seen in the final theoretical light curve at 
$t \sim 1.6$ d at all. The energy-injection model seems to be the 
most preferred choice. When an amount of energy that equals the initial 
kinetic energy of the GRB ejecta is added instantly into the blastwave 
at $t \sim 1.6$ d, a marked rebrightening emerges, which, although is 
still not rapid enough, gives an acceptable explanation to observations. 

Since the rebrightening of GRB 030329 is so rapid, we have to employ 
an instant energy-injection process in our calculation. In reality, 
this is most likely corresponding to the energy supplying process by 
a relatively slow shell which carries a significant amount of kinetic
energy but is ejected at a comparatively late stage by the central 
engine. Usually, the shell is very thin, with a width of $\sim 10^6$ 
--- $10^8$ cm, just as other more rapid shells that produce internal 
shocks and give birth to the main GRB. The shell moves 
outward at approximately a constant speed in a 
dilute environment that has been swept by previous shells. At the 
observer's time $t \sim 1.6$ d, when the shell finally catches up 
with the main blastwave, its thickness may increase slightly, but will 
still reasonably be much smaller than the radius of the blastwave. 
Interaction of this shell with the preceding blastwave can then be 
completed in a short time, producing an instant energy injection. 

For GRB 030329, Granot et al. (2003) suggested that there are 
a total of four energy-injection events within $\sim 9$ days after 
the burst trigger, which help to explain the observed variations 
between  $10^5$ s --- $10^6$ s. 
Here, when we extend the time span to $t \sim 80$ d, 
we identify a further energy injection event occurring at 
$1.2 \times 10^6$ s. In fact, similar multiple energy 
injections have also been suggested in another famous event,
GRB 021004, by de Ugarte Postigo et al. (2005). In that case, a total
of up to seven energy injections have been employed to explain the 
complex multiband afterglow light curves. 
In the standard fireball model of GRBs, afterglows
are deemed to largely lose their memory of the central engine. But 
the energy-injection shells are valuable fossils left by the central
engine. Careful study on these shells may provide important clues
for the central engine and the GRB trigger mechanism. 

However, GRB 030329 is a very complex event (Zeh, Klose, \& Kann 2005). 
Even in our best fit to the optical afterglow by engaging 
the energy-injection model (i.e., the solid line in Figure~3), 
there are still some obvious problems. The observed jet break 
at $t \sim 0.5$ d is not satisfactorily fitted; The theoretical 
rebrightening at $t \sim 1.6$ d is still not rapid enough; The 
observed subtle light curve variations during $2 \times 10^5$ s
--- $10^6$ s are not well accounted for. Solving these problems
may need the consideration of many further details, or even substantial 
revision of the model. Since the observational data are 
unprecedentedly prolific, GRB 030329 is undoubtedly a valuable 
sample. We suggest that a further complete, satisfactory fit to the R-band 
light curve (or even multi-band observations, ranging from radio to 
X-rays) should deserve trying, which will definitely provide useful 
information on the physics of GRBs and afterglows. 
 
Finally, we should bear in mind that the two-component jet model 
actually also has its own advantage when applied to GRB 030329: the emission
from the wide component can potentially give a natural explanation 
to the very late afterglow, but in the energy-injection model, a
further energy injection process will have to be assumed for the afterglow 
beyond $10^6$ s, which is somewhat artificial and makes us uncomfortable. 
The two-component jet is still a possibility, although it cannot be used 
to explain the rebrightening episodes. In fact, it is probable that a 
compound model may be taking effect in the case of GRB 030329. The event
may basically be due to a two-component jet, with the narrow component 
accounting for the early afterglow ($t \leq 1.0 \times 10^5$ s) and the 
wide component accounting for the late afterglow ($t \geq 3.0 \times 
10^5$ s). At the same time, an additional energy injection may happen 
to the narrow component at about 1.6 d, which explains the observed 
rebrightening. It is interesting to note that another novel idea that 
may possibly reconcile the two-component jet model and the 
energy-injection process has also been proposed by Resmi et al. (2005) 
recently. They suggested that there might be only one narrow jet 
initially in GRB 030329. But at a time around or before $\sim 1.5$ d, a
possible re-energization event may take place, refreshing the initial
narrow jet into a second, ``wide'' jet. They noted that the half 
opening-angle of the initial narrow jet, due to side expansion, has 
already increased to a value which equals that required for the wide 
jet. While the scenario itself is also plausible, the
detailed physical process still needs to be clarified extensively.

\acknowledgments

We thank the anonymous referee for many valuable comments that lead 
to an overall improvement of this study and especially for helping us
to access the observational data of GRB 030329.  
This research was supported by a RGC grant of the Hong Kong Government, 
and also partly supported by the Special Funds for Major State
Basic Research Projects, the National Natural Science Foundation
of China (Grants 10233010, and 10221001), and the Foundation 
for the Author of National Excellent Doctoral Dissertation of P. R. 
China (Project No: 200125).

\clearpage



\begin{figure}
\epsscale{1.05}
\plotone{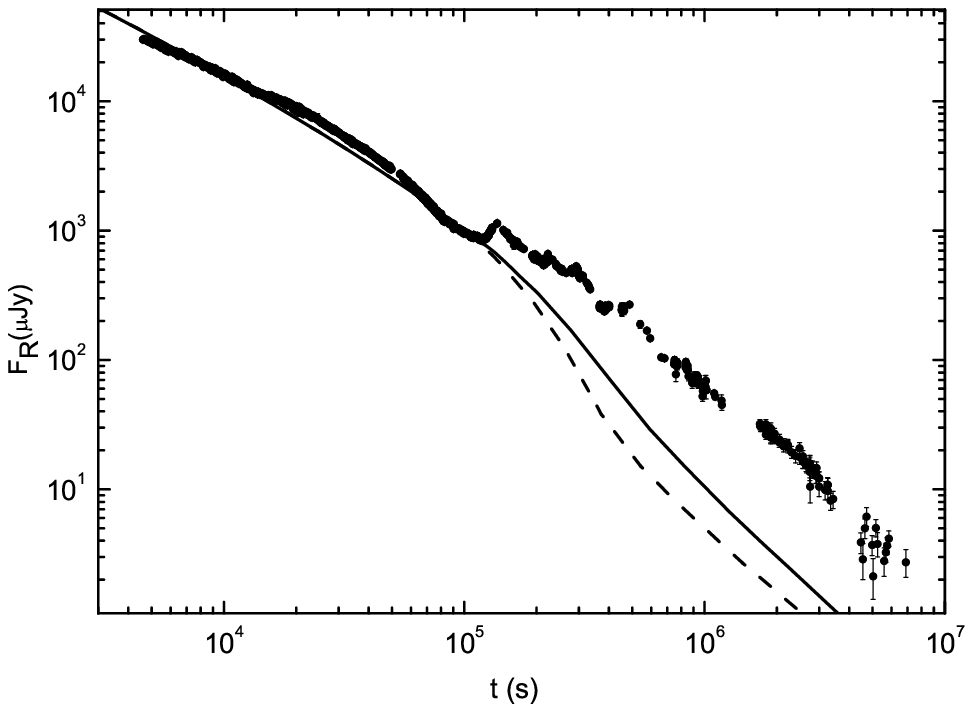}
\caption{An illustration of our fit to the R-band afterglow light curve of 
GRB 030329 by using the density-jump model. Observed points correspond to  
pure afterglow emission, derived by subtracting the host galaxy contribution 
and the supernova contribution from the data of Lipkin et al. (2004). 
In the solid line, the number
density of the circumburst medium is assumed to increase abruptly from 
2 cm$^{-3}$ to 20 cm$^{-3}$ at the observer's time $t = 1.1 \times 10^5$ s. 
In the dashed line, the density increases from 2 cm$^{-3}$ to 200 cm$^{-3}$ 
at the same time. Other parameters involved in this figure have been given
in \S 3.1 . In both cases, the model fails to reproduce the observed 
rebrightening at $t \sim 1.6$ d and the observed flux excess 
thereafter. 
\label{fig1}}
\end{figure}

\clearpage

\begin{figure}
\epsscale{1.15}
\plottwo{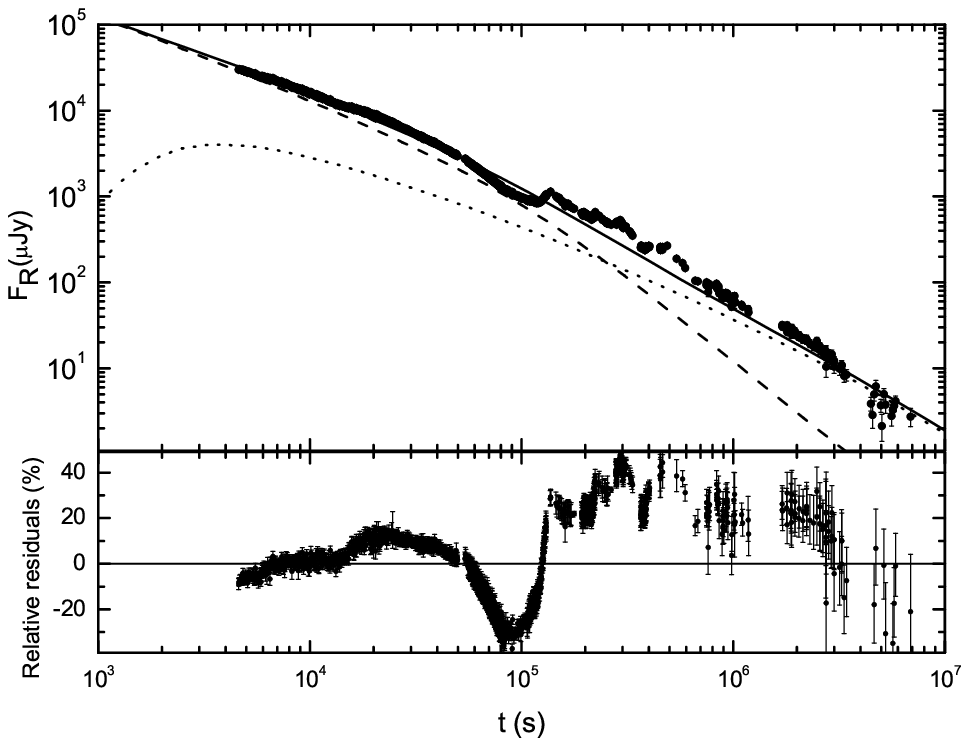}{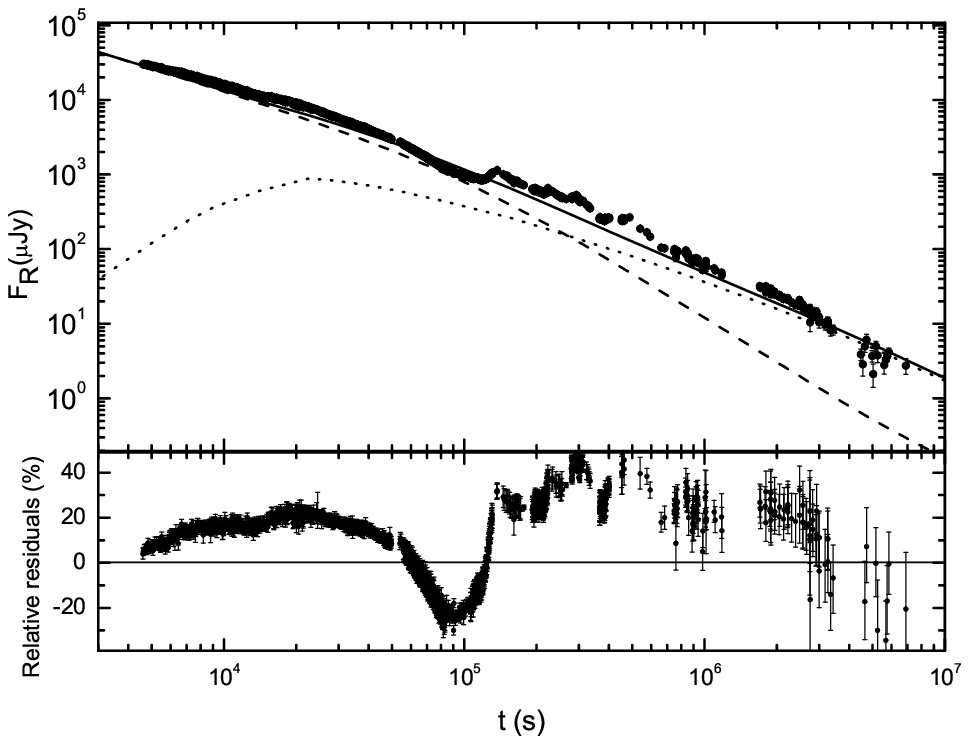}
\caption{(a) Our best fit to the R-band afterglow light curve of GRB 
030329 using the two-component jet model. Observed points correspond to 
pure afterglow emission, derived from the data of 
Lipkin et al. (2004). The dashed line corresponds 
to emission from the narrow component, the dotted line corresponds to 
emission from the wide component, and the solid line is the total light curve. 
In the lower panel, the relative residual of the fit is plotted. 
Parameters involved in this figure have been given in \S 3.2 .
(b) Same as (a), except that the wide component 
is now assumed to be a hollow cone since its central portion is 
occupied by the narrow component. Note that in both (a) and (b), no 
obvious rebrightening can be seen in the modeled light curves at 
$t \sim 1.6$ d, thus the model is not preferred by observations. 
\label{fig2}}
\end{figure}

\clearpage

\begin{figure}
\epsscale{1.05}
\plotone{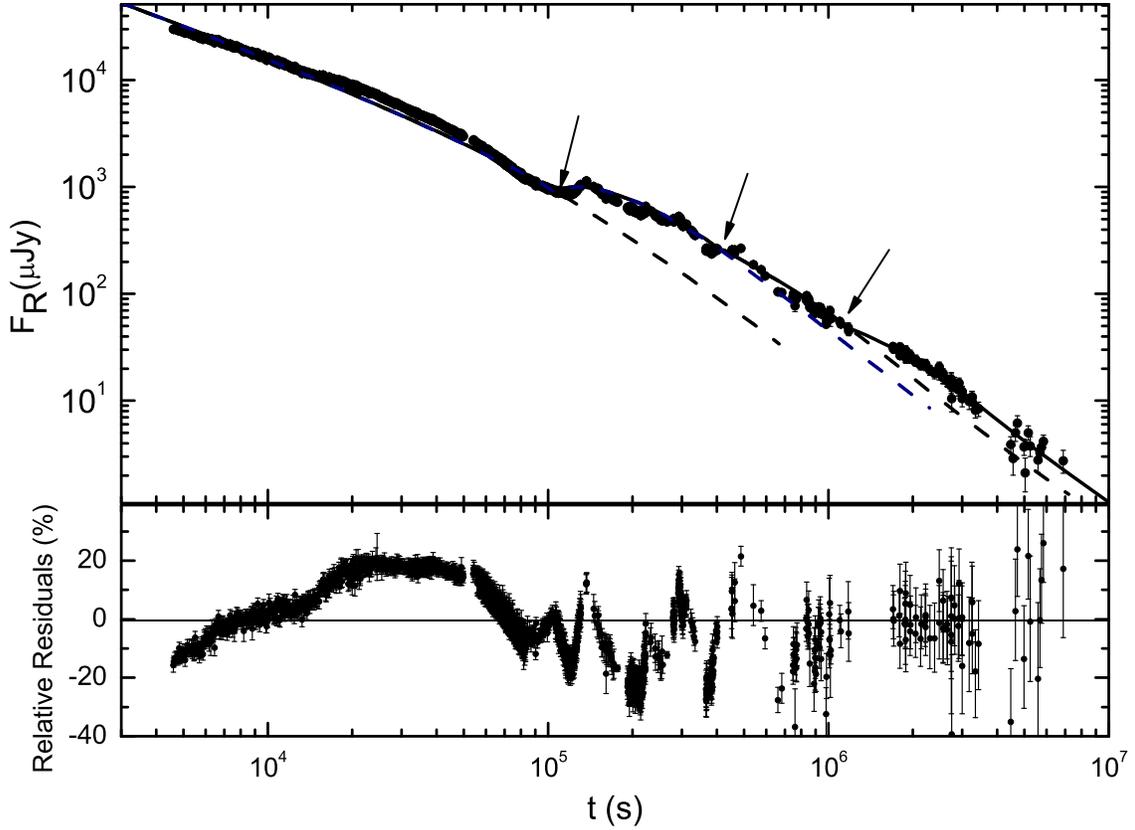}
\caption{Our fit to the R-band afterglow light curve of GRB 
030329 in light of the energy-injection model. Observed points 
correspond to pure afterglow emission, derived from the data 
of Lipkin et al. (2004). 
The solid line corresponds to our best fit by engaging 3 energy
injections, which occur at $t = 1.1 \times 10^5$ s, 
$4.0 \times 10^5$ s, and $1.2 \times 10^6$ s, as marked by 
arrows. The injected energies are 1.0 $E_0$, 0.4 $E_0$, 
and 0.6 $E_0$, respectively.  
For comparison, the dashed lines illustrate the theoretical afterglows 
when no further energy is supplied. 
Parameters involved in this figure are the same as those in Figure~1. 
In the lower panel, the relative residual of the solid line is plotted. 
\label{fig3}}
\end{figure}



\end{document}